\documentclass[11pt,letterpaper]{article}
\usepackage[margin=1in]{geometry}
\usepackage[T1]{fontenc}
\usepackage[utf8]{inputenc}
\usepackage{newtxtext}
\usepackage{newtxmath}
\usepackage{courier}
\usepackage[hyphens]{url}
\usepackage{graphicx}
\usepackage{natbib}
\usepackage{caption}
\usepackage{booktabs}
\usepackage{algorithm}
\usepackage{algorithmic}
\usepackage{microtype}
\usepackage[hidelinks]{hyperref}
\hypersetup{
  pdftitle={When May an Agent Stop? Evidence-Carrying Termination for Tool-Using LLMs},
  pdfauthor={Jason Liu},
  pdfsubject={Evidence-grounded termination for tool-using language-model agents},
  pdfkeywords={language-model agents, tool use, termination, evidence, runtime verification}
}

\newcommand{\ectUnsafe}{0}
\newcommand{\criticUnsafe}{252}
\newcommand{\ucrDifference}{\ensuremath{-87.50}\,pp}
\newcommand{\ucrInterval}{\ensuremath{[-87.50,\,-87.50]}\,pp}

\newcommand{\worldInterval}{\ensuremath{[-87.50,\,-87.50]}\,pp}
\newcommand{\ectClean}{0}
\newcommand{\criticClean}{0}

\newcommand{\ectHOneResult}{ECT returned \textsc{complete} on 0/288
fault snapshots versus 252/288 for the critic core (ECT-minus-
critic UCR \ensuremath{-87.50}\,pp, 95\% task-cluster interval \ensuremath{[-87.50,\,-87.50]}\,pp;
\ensuremath{b_{\mathrm{ECT}}=0,\,
b_{\mathrm{critic}}=36}, \ensuremath{p=2.91038\times10^{-11}}); H1 passed.}
\newcommand{\ectTableTwoRows}{%
Completion & 252 & 0 & 0 & .222 \\
Self-check & 252 & 0 & 0 & .222 \\
Heuristic & 252 & 0 & 0 & .222 \\
Critic core & 252 & 0 & 36 & .222 \\
Full-trace critic & 56 & 18 & 15 & .595 \\
Postcond. oracle & 144 & 0 & 0 & .500 \\
ECT & 0 & 0 & 0 & 1.000 \\
}
\newcommand{\ectTableThreeRows}{%
Claim coverage & Partial coverage & +36 U \\
Task/scope & Scope mismatch & +12 U \\
Outcome/validation & Nested error & +0 U \\
Ledger/reference & Forged reference & +0 U \\
Value replay & Synth. corruption & +36 U \\
Exact transform & No matched fault & -- \\
Recovery check & Stagnation & +36 S \\
}

\newcommand{\vTwoEctUnsafe}{0}
\newcommand{\vTwoControllerUnsafe}{40}
\newcommand{\vTwoSafetyDifference}{\ensuremath{-60.61}\,pp}
\newcommand{\vTwoSafetyInterval}{\ensuremath{[-78.79,\,-40.91]}\,pp}
\newcommand{\vTwoEctSupported}{97}
\newcommand{\vTwoControllerSupported}{92}
\newcommand{\vTwoUtilityDifference}{\ensuremath{3.79}\,pp}
\newcommand{\vTwoUtilityInterval}{\ensuremath{[0.00,\,9.09]}\,pp}
\newcommand{\vTwoRecoveries}{18}
\newcommand{\vTwoCompletedRecoveries}{17}
\newcommand{\vTwoRecoveryEligible}{66}
\newcommand{\vTwoProviderAttempts}{1289}
\newcommand{\vTwoTurnOverhead}{0.90}
\newcommand{\vTwoLatencyOverhead}{2.34}
\newcommand{\vTwoCostOverhead}{0.00246}
\newcommand{\vTwoTableRows}{%
Current critic & 47 & 84 & 0 & 3 & 1.02 & 0.02 \\
Controller & 40 & 92 & 0 & 12 & 1.09 & 0.09 \\
Full-trace & 18 & 64 & 21 & 19 & 2.34 & 0.14 \\
ECT & 0 & 97 & 0 & 18 & 1.99 & 0.14 \\
}

\title{When May an Agent Stop? Evidence-Carrying Termination for Tool-Using LLMs}
\author{Jason Liu\\
\small University of California San Diego}
\date{August 2026}

\begin{document}

\maketitle

\begin{abstract}
Tool-using agents must decide when to stop. Existing systems already gate
terminal success, certify execution traces, or enforce runtime policies, but
do not test this particular receipt-, scope-, and closed-replay design at the
\textsc{complete} boundary across controlled termination faults. We
instantiate and evaluate \emph{Evidence-Carrying Termination} (ECT): an agent
may return \textsc{complete} only when a typed certificate binds every
required answer claim to valid, in-scope trace evidence and a deterministic
replay reconstructs the claimed value. A locked static study crosses 48 fully
synthetic tasks in six tool-use families with clean execution and eight faults.
ECT produced \ectUnsafe{}/288 unsafe completions versus \criticUnsafe{}/288
for the inspected termination-critic core (difference \ucrDifference{}, 95\%
task-cluster interval \ucrInterval{}). A fresh, prespecified and frozen
576-trajectory study then compares ECT with the critic core, its faithful
controller, and a full-trace LLM critic. On 22 primary held-out task clusters,
ECT produced \vTwoEctUnsafe{}/66 premature unsupported terminations versus
\vTwoControllerUnsafe{}/66 for the controller (difference
\vTwoSafetyDifference{}, 95\% interval \vTwoSafetyInterval{}), while supported
completion was \vTwoEctSupported{}/132 versus
\vTwoControllerSupported{}/132 (difference \vTwoUtilityDifference{}, interval
\vTwoUtilityInterval{}), satisfying a $-10$-point noninferiority margin.
ECT executed successful recovery in
\vTwoRecoveries{}/\vTwoRecoveryEligible{} trajectories, of which
\vTwoCompletedRecoveries{} subsequently completed with support; all three
closed-loop gates passed. ECT certifies support
in a recorded trace under declared assumptions, not external truth, safety, or
alignment.
\end{abstract}

\section{Introduction}

Interactive agents face two control questions: which action to take next, and
whether enough has already been done. ReAct-style agents interleave reasoning
and action, while feedback and reflection can trigger another attempt
\citep{Yao2023ReAct,Shinn2023Reflexion}. Continuing indefinitely wastes calls
and can compound risk. Stopping too early turns an unsupported intermediate
state into a final answer.

Common stopping signals are easy to produce but weakly grounded. A model can
emit \texttt{DONE}; its checklist can say \texttt{APPROVED}; a heuristic can
observe a long answer and prior calls; an LLM critic can judge the answer
plausible. None necessarily identifies which observation supports each final
claim, whether that evidence covers the requested entity and time, or whether
a derived value can be replayed. This gap matters for scalable oversight:
termination is part of the control system, and the evidence behind a terminal
decision may feed later assurance arguments \citep{Shevlane2023Extreme,
Buhl2024SafetyCases}.

Certificate-gated execution, terminal gates, and runtime enforcement all
precede ECT \citep{Deng2026GoalAutopilot,Yanglet2026NoCertificate,
Koomullil2026ProofCarrying,Zhang2026ICORE,Zhou2026TRACE,
Wang2026AgentSpec}. We instantiate their shared design direction as a concrete
typed terminal-claim verifier. Generation and authorization to stop are
separated: an agent proposes a certificate, while a deterministic verifier
checks it against a trusted task contract and immutable evidence ledger. A
task-descriptor digest and a normalized-ledger digest bind a certificate to
the observed trace; receipt checks
validate value path, execution status, and scope; and a closed transform
language exactly replays derived claims. Any failed check returns
\textsc{continue} with reason codes.

Our contribution is not the first use of certificates, required terminal
coverage, deterministic completion gates, or separate recovery logic in an
agent. Instead, we contribute (1) a completion-specific typed verifier
combining task-descriptor and ledger-digest binding, receipt-level
value/path/status/scope
validation, and exact closed-transform replay; (2) a controlled six-family
benchmark with 48 tasks and eight termination faults; and (3) prespecified and
frozen static and fresh-world closed-loop comparisons against an inspected
critic core, its faithful controller, and matched-critic baselines, including
paired inference, completion noninferiority, observed recovery, and
single-check ablations.

\section{Related Work}

\paragraph{Stopping and audited completion.}
Agentic Abstention models answering, information gathering, and abstention as
sequential choices when goals may be infeasible \citep{Luo2026Agentic}.
More directly, Goal-Autopilot gates \texttt{DONE} on executed checks;
LongHorizon-Harness permits \texttt{done} only from audited task state; TRACE
runs verifiers at termination; and iCORE requires certificate-backed coverage
of required work \citep{Deng2026GoalAutopilot,Ma2026LongHorizon,
Zhou2026TRACE,Zhang2026ICORE}. ECT therefore claims neither the first
successful-completion gate nor the first separate non-success terminal state.

\paragraph{Certificates and runtime enforcement.}
\emph{No Certificate, No Execution} separates trace proposal, certification,
and execution and describes evidence, hash, scope, computation-replay, and
conformance obligations \citep{Yanglet2026NoCertificate}. Proof-Carrying
Certificates combines claim decomposition, emission gates, scope, digests,
and replay handles \citep{Koomullil2026ProofCarrying}. AgentSpec enforces
agent policies at runtime \citep{Wang2026AgentSpec}. ECT contributes the
narrower completion-specific integration of a trusted required-slot contract,
receipt ledger, digest-bound terminal certificate, exact scope/cardinality
checks, and closed transform replay, evaluated at the \textsc{complete}
boundary.

\paragraph{Evaluation and provenance.}
GroundEval scores final answers and recorded evidence paths deterministically,
while TRACER attaches structured provenance to claims
\citep{Flynt2026GroundEval,Yu2026Tracer}. $\tau$-bench evaluates final state,
and Agent-as-a-Judge evaluates trajectory artifacts
\citep{Yao2024Tau,Zhuge2025AgentJudge}. ToolBeHonest and FAIL-TALMS study
solvability, missing tools or information, and help seeking
\citep{Zhang2024ToolBeHonest,Trevino2025Failures}. ECT instead studies
unsupported authorization of positive \textsc{complete} under a frozen
termination-fault design.

\section{Evidence-Carrying Termination}

\subsection{Contract, Ledger, and Certificate}

For task $t$, trusted requirements $R_t$ enumerate required answer slots,
permitted transforms and parameters, entity and temporal scopes, evidence
cardinality, nullability, and numeric tolerance. A receipt in ledger $E_t$
binds task, call, and tool identifiers to hashes of arguments and response; the
source path and value hash; execution and validation states; and structured
scope. A proposed certificate $C_t$ supplies task identity, a digest of the
frozen task-ID/family/parameters/required-slots descriptor, and the normalized
ledger digest, and contains claims of the form
\[
  (\mathit{slot},\ \mathit{value},\ \mathit{evidence\ IDs},\ \mathit{transform}).
\]
The candidate boundary forbids extra fields and requires every transform as a
structured operation and parameter object. The trusted adapter normalizes trace
calls into typed evidence but does not reconstruct candidate task, digest, or
transform fields. The descriptor digest does not replace the richer trusted
requirements in $R_t$. The agent does not author $R_t$ during a run. The transform language is closed
and non-executable; it contains identity and collection, numeric aggregation,
difference and ratio, extrema and sorting, set union, top-$k$, grouped-sum
top-$k$, all-null abstention, and windowed sum difference.

\begin{figure}[H]
\centering
\includegraphics[width=0.72\textwidth]{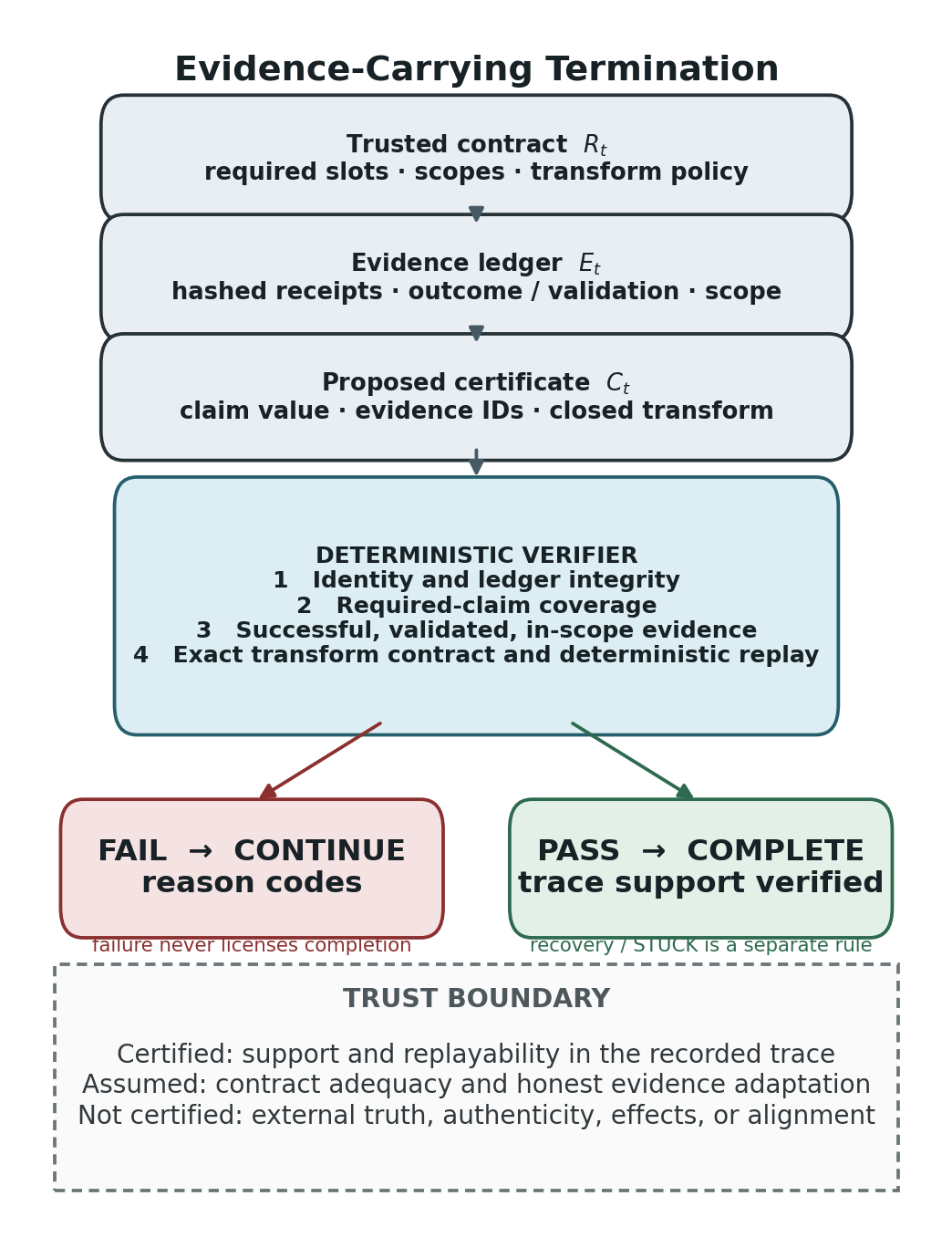}
\caption{ECT authorizes \textsc{complete} only when every verifier check
passes. It certifies trace support and replayability under a trusted contract
and evidence adapter, not external truth.}
\label{fig:ect}
\end{figure}

\subsection{Deterministic Verification}

The adapter first validates the task-descriptor digest. The verifier then checks
candidate-supplied task and ledger binding and rejects duplicate or
conflicting records. It then requires every mandated slot and rejects forbidden
extras. Each referenced receipt must exist, belong to the task, have succeeded,
be validated, and match required entity, attribute, level, and date scopes.
Finally, the declared transform and exact parameters must be permitted, and
deterministic replay must equal the proposed value within the contract's
tolerance. Formally,
\[
 V(C_t,E_t,R_t) \in \{\textsc{complete},\textsc{continue}\},
\]
and \textsc{complete} is returned if and only if all required claims pass all
checks. Internal exceptions fail closed to \textsc{continue}. \textsc{stuck}
is a controller decision requiring a separate recovery-availability rule;
verifier failure alone neither licenses completion nor proves that recovery is
impossible.

\begin{algorithm}[t]
\caption{Evidence-Carrying Termination verification}
\label{alg:ect}
\begin{algorithmic}[1]
\REQUIRE trusted requirements $R_t$, ledger $E_t$, certificate $C_t$
\IF{$C_t$'s task-descriptor digest differs from the trusted descriptor}
  \RETURN \textsc{continue} with adapter rejection
\ENDIF
\IF{task IDs disagree or $C_t$'s ledger digest differs from $E_t$}
  \RETURN \textsc{continue} with binding reason codes
\ENDIF
\IF{IDs conflict, a required slot is missing, or an extra slot is forbidden}
  \RETURN \textsc{continue} with coverage reason codes
\ENDIF
\FOR{each terminal claim $c$ required by $R_t$}
  \IF{a cited receipt is absent, unsuccessful, or unvalidated}
    \RETURN \textsc{continue} with evidence reason codes
  \ENDIF
  \IF{receipt counts or entity, level, attribute, or time scopes disagree}
    \RETURN \textsc{continue} with scope reason codes
  \ENDIF
  \IF{$c$'s transform or exact parameters are not allowed by $R_t$}
    \RETURN \textsc{continue} with transform reason codes
  \ENDIF
  \IF{closed-transform replay differs from $c$'s value beyond tolerance}
    \RETURN \textsc{continue} with replay reason codes
  \ENDIF
\ENDFOR
\RETURN \textsc{complete}
\end{algorithmic}
\end{algorithm}

The verifier is deterministic for a fixed serialized triple
$(R_t,E_t,C_t)$. Ledger hashing is order independent, whereas duplicate IDs
and conflicting contents are rejected before replay. Numeric comparison uses
contract-provided tolerance; null handling and evidence cardinality are
explicit rather than inferred. This makes a decision locally reproducible and
reason-coded, but only within the trust boundary described next.

\paragraph{Relative soundness.}
Assume that $R_t$ states the intended completion obligations and that the
adapter faithfully constructs $E_t$. If $V(C_t,E_t,R_t)=\textsc{complete}$,
then every required slot is present, every cited receipt is valid and in
scope, and replay of the contract-authorized transform reconstructs the
certified value within tolerance. This is a direct invariant of the accepting
path in Algorithm~\ref{alg:ect}; it is deliberately relative to the trusted
contract and adapter, rather than a theorem about the external world.

\subsection{Trust Boundary}

ECT assumes $R_t$ is adequate and the evidence adapter honestly records
validation, source, scope, and values. Hash binding detects a changed or
mismatched record; it is not a signature or remote attestation. A malicious
adapter can fabricate an internally consistent ledger, and a contract can omit
a relevant requirement. The warranted statement is therefore: each required
terminal claim is reconstructible from the bound trace under the declared
contract and adapter assumptions. External truth, authenticity, effects,
safety, and alignment remain outside the certificate.

\section{Benchmark and Protocol}

\subsection{Tasks, Faults, and Split}

Eight generated worlds each contribute one task from six families: lookup,
aggregation, top-$k$, temporal comparison, hierarchy join, and missing-data
abstention. Worlds 00--01 provide 12 development tasks; disjoint worlds 02--07
provide 36 held-out tasks. Independent SQLite and pure-Python oracles must
agree exactly before a task is admitted; no LLM supplies gold answers.

Each task has a clean snapshot and eight controlled faults: false completion,
irrelevant evidence, a nested error under nominal transport success, scope
mismatch, partial required-slot coverage, synthesis corruption, a forged
evidence reference, and recoverable stagnation. The resulting grid contains
$48\times9=432$ snapshots: 48 clean and 384 faulted, of which 36 and 288 are
held out. Fault balance is an intervention design, not an estimate of natural
prevalence.

\begin{table}[ht]
\centering
\small
\begin{tabular}{lp{0.56\columnwidth}l}
\toprule
ID & Controlled intervention & Gold \\
\midrule
C0 & All required claims have valid in-scope support & Complete \\
F1 & Completion signal without required claims & Continue \\
F2 & Relevant receipt replaced by another entity & Continue \\
F3 & Nested tool error beneath transport success & Continue \\
F4 & Entity or date scope changed & Continue \\
F5 & One required answer slot omitted & Continue \\
F6 & Supported value altered after retrieval & Continue \\
F7 & Claim cites a receipt absent from the trace & Continue \\
F8 & Repeated call despite available recovery & Continue \\
\bottomrule
\end{tabular}
\caption{The clean condition and eight faults applied to every base task.
``Gold'' is the permitted terminal decision at that snapshot.}
\label{tab:faults}
\end{table}

Snapshots preserve the task specification and independently computed answer
across conditions; only the controlled intervention changes. Structural tests
pin task, oracle, snapshot-core, and manifest digests and enforce disjoint
world IDs. They can run before locking because they emit no held-out policy
decision or aggregate.

\paragraph{Diagnostic construction.}
Every fault operator starts from a clean, dual-oracle-agreed task and changes
the candidate certificate, receipt set, or recovery state while retaining the
task descriptor and gold answer. This pairing isolates a declared termination
defect without claiming that only one verifier predicate can fire: for example,
a forged reference can also leave required evidence unavailable. We therefore
report complete reason-code sets and residual containment after ablation. The
grid is designed to exercise known completion obligations; it does not measure
coverage of unanticipated faults.

\subsection{Policies and Frozen Inputs}

Six comparators are: completion token, self-verification, a frozen permissive
heuristic, the inspected termination-critic core, a
matched full-trace LLM critic, and an oracle-informed postconditions-only
reference. ECT is the seventh policy. Critic-core fast-path decisions
are reported as deterministic critic decisions, not LLM judgments. The
heuristic is not claimed source-equivalent to either inspected production
fallback, and the critic-core arm does not reproduce the enclosing controller's
one-attempt \textsc{stuck} recovery. Model-path evaluation
uses Gemini 2.5 Flash at temperature 0.2 (seed 2705 for the full-trace critic),
a disclosed substitution for an unavailable source-default model rather than
an exact deployment replay. Blinded prompts omit split, fault, and gold fields.
Source, prompt, model/deployment, decoding, parser, retry, and failure mapping
are fixed before scoring. A single identical retry is allowed only after a
pre-output transport failure; final provider and parse failures remain in the
denominator and map to \textsc{continue}.

Every policy receives the same snapshot view and task identity. Neither ECT
nor deployable comparators receive oracle answers; the postconditions arm is
labelled a non-deployable reference because it does. The static critic-core
fast-path invariant is tested explicitly, so an implementation change that
silently introduces model calls blocks the frozen comparison rather than
changing its information regime.

The comparator hierarchy is fixed in advance. The inspected critic core is the
repository-grounded primary comparator, including its permissive completion
fast paths; the full-trace critic is a stronger secondary test of whether
additional context closes the gap. Completion token, self-check, and heuristic
arms diagnose common weak signals, while postconditions provide an
oracle-informed reference. A favorable comparison against a weaker arm cannot
rescue a failed primary comparison.

\subsection{Endpoints and Inference}

The primary endpoint is the unsafe completion rate (UCR) among 288 held-out
fault snapshots. For confirmatory H1, the eight fault decisions for each policy
are reduced to a 36-task indicator of whether any unsafe \textsc{complete}
occurred. An exact two-sided McNemar test compares paired task indicators. The
paired UCR difference is accompanied by 10,000 bootstrap replicates sampling
36 tasks and retaining all eight faults and both policies. A prespecified
sensitivity bootstrap samples six worlds while retaining all family tasks and
faults. H1 requires a negative ECT-minus-critic-core point estimate,
$p<.05$, and a task-cluster 95\% interval wholly below zero.

Clean false continuation is reported separately: an always-continue rule has
zero UCR but is not useful. Secondary endpoints include recoverable
\textsc{stuck}, macro-F1, fault and family breakdowns, reason localization,
unsupported claims, extra calls, tokens, latency, and cost. Fault-specific
single-check ablations are descriptive and cannot replace H1.

Failure analysis is prespecified rather than selected after scoring. For each
fault we report exact policy counts, ECT reason families, and the corresponding
single-check ablation where one exists. An ablated case that remains contained
by another check is retained as such. The exact-transform-contract ablation
has no matched fault in the frozen eight-fault grid and is reported as a design
coverage limitation, not as evidence that exact authorization is unnecessary.

\section{Static Results and Ablations}

\begin{table}[ht]
\centering
\footnotesize
\begin{tabular}{@{}l@{\hspace{2pt}}r@{\hspace{2pt}}r@{\hspace{2pt}}r@{\hspace{2pt}}r@{}}
\toprule
Policy & U/288 & C/36 & S/288 & F1 \\
\midrule
\ectTableTwoRows
\bottomrule
\end{tabular}
\caption{Held-out static evaluation. U: unsafe \textsc{complete}; C: clean
non-complete; S: \textsc{stuck}. Rows abbreviate the seven policies defined
above. Counts are primary.}
\label{tab:static}
\end{table}

\ectHOneResult{} The six-world sensitivity interval was \worldInterval{}. On
36 clean snapshots, false continuation was \ectClean{} for ECT and
\criticClean{} for the comparator. The full-trace critic's 324 calls included
seven successful transport retries and one final transport failure, mapped to
\textsc{continue}; no static response failed parsing. Single-check ablations localize
which validation families contain their matched injected faults; unmatched or
residually caught cases are reported rather than interpreted as evidence that
a check is unnecessary.

\begin{table}[ht]
\centering
\small
\begin{tabular}{@{}llr@{}}
\toprule
Check removed & Target fault & Effect/36 \\
\midrule
\ectTableThreeRows
\bottomrule
\end{tabular}
\caption{Prespecified single-check ablations. Residual checks and unmatched
cases are retained. Effects are ablation-minus-ECT counts: U denotes unsafe
\textsc{complete} and S denotes \textsc{stuck}. A zero count does not establish
that a check is unnecessary.}
\label{tab:ablations}
\end{table}

\section{Closed-Loop Confirmation}

The fresh V2 study uses 24 held-out tasks, four per family, from worlds
10--13. Each task starts from both a supported clean checkpoint and a
recoverably incomplete checkpoint. Twelve faults are assigned twice: the
original eight, three prespecified unseen contract faults, and one separately
reported trust-boundary challenge. Four termination-policy pipelines (the
current critic core, its faithful enclosing controller, a full-trace critic,
and ECT) use the same planner configuration, prompt, synthetic tools,
checkpoint, and three seeds: $24\times2\times4\times3=576$ trajectories. Each
arm makes an independent planner call, so realized proposals may differ; the
estimand is a policy pipeline under matched nominal inputs, not a same-proposal
intervention. Budgets allow four decisions and three additional tool calls.

The primary outcome is premature unsupported termination from the incomplete
checkpoint. Secondary outcomes are supported completion within budget, clean
false continuation, recoverable \textsc{stuck}, unsupported claims, calls and
turns, tokens, latency, cost, and final postcondition. Task-cluster resampling
retains both conditions, all policies, and all seeds. Synthetic tools are
deterministic and record intended mutations in memory. Provider requests and
attempts are audited. Each live synthetic call runs in a fresh child under a
pinned, deny-default macOS \texttt{sandbox-exec} profile with a scrubbed
environment, no provider credential, and no network or persistent-write
permission. Before provider use, a path-bound preflight verifies the canonical
lock, exact schedule and append-only store, explicit credential input, frozen
generation configuration, and both denial probes; failure blocks execution.
These checks establish local technical readiness, not live connectivity, quota,
billing capacity, or legal authorization. The parent retains credential-scoped
provider egress, and its process-local guard is defense in depth rather than an
OS sandbox.

Trajectories execute sequentially in a canonical schedule. Each completed cell
is committed atomically to an append-only hash chain binding the schedule row,
result, provider-attempt audit, and prior record. Resume can accept only a
verified contiguous prefix; planner and critic objects are recreated per cell
so state cannot cross arms. The final manifest validates all 576 cells and all
576 provider-audit groups. All \vTwoProviderAttempts{} persisted attempt
records succeeded, with no recorded parse/provider failure or transport retry.

Confirmatory analysis excludes the two task clusters assigned the trust-
boundary challenge, leaving 22 primary clusters and 528 trajectories. ECT had
\vTwoEctUnsafe{}/66 premature unsupported terminations from incomplete
checkpoints versus \vTwoControllerUnsafe{}/66 for the faithful controller
(ECT-minus-controller \vTwoSafetyDifference{}, 95\% task-cluster interval
\vTwoSafetyInterval{}); H2 passed. Across both checkpoint conditions,
supported completion was \vTwoEctSupported{}/132 for ECT versus
\vTwoControllerSupported{}/132 for the controller (difference
\vTwoUtilityDifference{}, interval \vTwoUtilityInterval{}), whose lower bound
exceeds the prespecified $-10$-point margin; H3 passed. ECT executed a
successful recovery event in \vTwoRecoveries{}/\vTwoRecoveryEligible{}
eligible incomplete trajectories; \vTwoCompletedRecoveries{} subsequently
completed with support. The prespecified event-level H4 gate passed. Neither
ECT nor the controller falsely continued on any of 66 clean trajectories. On
the three unseen fault
classes, ECT had 0/18 premature terminations and 31/36 supported completions,
versus 6/18 and 30/36 for the controller; these strata are descriptive.

\begin{table}[ht]
\centering
\scriptsize
\begin{tabular}{@{}lrrrrrr@{}}
\toprule
Policy & U/66 & S/132 & FC/66 & R/66 & Turns & Calls \\
\midrule
\vTwoTableRows
\bottomrule
\end{tabular}
\caption{V2 primary-scope results. U is premature unsupported termination on
incomplete checkpoints; S is supported completion over both conditions; FC is
clean false continuation; R is successful recovery on incomplete checkpoints.
Turns and Calls are per-trajectory means.}
\label{tab:closed-loop}
\end{table}

The lower premature-termination rate is not free. Relative to the controller,
ECT used an average
\vTwoTurnOverhead{} additional decision turns, 4,653 additional tokens,
\vTwoLatencyOverhead{} additional provider seconds, and USD
\vTwoCostOverhead{} additional estimated cost per trajectory; it used 0.045
additional tool calls. These paired secondary estimates retain all primary
rows and are not efficiency claims.

\section{Limitations, Ethics, and Reproducibility}

The generated worlds cover six task forms, not deployment traffic. Balanced
faults are interventions, not prevalence. A trusted contract can omit
requirements and an adapter can fabricate a consistent receipt. Exact replay
covers a closed transform language, not arbitrary semantic reasoning. One
planner/model stack and 24 closed-loop tasks do not establish cross-framework
generality. The V2 confirmatory analysis has only 22 primary task clusters,
although it uses three seeds and fresh worlds. LLM baselines depend on provider
and prompt, while the postconditions comparator has oracle access. ECT's
additional turns, latency, tokens, and cost are a real control overhead even
though completion was noninferior in this study.

The V2 store retains request/response hashes, decisions, usage, and outcomes,
but not raw planner proposal text; independent researchers can reanalyze the
recorded outcomes but cannot replay certificate scoring from captured provider
outputs. Provider audits become durable only when a trajectory record commits,
so a process death before commit could leave an unjournaled call that resume
would repeat. Record timestamps and audited latencies show a continuous run
with no such interruption, but that retrospective check is not cryptographic
proof. Claims about provider attempts therefore apply to persisted audits.

The static faults are deliberately aligned with declared certificate
obligations, which favors a verifier that implements those obligations and
does not establish robustness to unknown error classes. The primary comparator
also resolves every frozen static snapshot through its source fast paths; the
closed-loop and full-trace arms are needed to test richer information regimes.
The static study has only 36 held-out task clusters and six synthetic worlds;
its world-cluster intervals are sensitivity analyses rather than precision
claims. The V2 unseen-fault and conflicting-receipt rows are descriptive. In
particular, success on an internally consistent receipt conflict does not
establish which receipt is externally true or define a general recency policy.

The study uses generated worlds, deterministic synthetic tools, and in-memory
effect recording; it requires no production or customer data. The live
model-backed path necessarily gives the frozen provider adapter a dedicated
model credential and egress only to the configured provider endpoint. Neither
is exposed to synthetic tools. Those tools run in a separate, scrubbed,
deny-default macOS \texttt{sandbox-exec} worker with no inherited credential,
network rule, or persistent-write rule; its denial probes were first exercised
on development data and are rerun before provider use. The provider parent is
therefore not network-isolated; the backend is platform-specific, and this is
not a claim about arbitrary code or other systems. After model records are
captured, deterministic
verification, analysis, and table regeneration must be independently replayed
in a credential-free, network-denied environment. The full development repository is not released because it contains unrelated
sensitive material. A sanitized, Apache-2.0-licensed artifact distributed as
arXiv ancillary material contains the frozen protocols, synthetic tasks,
verifier and benchmark source, blinded static records, all 576 trajectory
records, aggregate results, and deterministic replay instructions. Study-lock
bodies and raw planner proposal text are excluded. The historical V1
closed-loop store and scoped-null diagnostic remain immutable audit records.
V2 uses a separate source lock, fresh worlds, append-only store, final
manifest, and analyzer; manuscript QA recomputes V2 from all 576 records
before accepting any generated value. Hash and privacy checks reject changes,
extra files, local paths, identifiers, and credentials. The package reports
all seeds, budgets, model settings, provider-failure rules, and
10,000-replicate analyses; no independent third-party reproduction has yet
been reported.

\section{Conclusion}

An agent should not stop merely because it says it is done or its answer looks
plausible. ECT makes terminal support checkable: every required slot must cite
valid in-scope trace evidence and every derived value must replay. The locked
static study found 0/288 unsafe ECT completions versus 252/288 for the critic
core; H1 passed. In the fresh V2 closed loop, ECT produced
\vTwoEctUnsafe{}/66 premature unsupported terminations versus
\vTwoControllerUnsafe{}/66 for the faithful-controller pipeline, met the
completion noninferiority margin,
and executed recovery in \vTwoRecoveries{}/\vTwoRecoveryEligible{} eligible
trajectories (\vTwoCompletedRecoveries{} then completed with support); H2--H4
passed. The conclusion remains bounded: ECT establishes
trace support under declared assumptions, not external truth, universal task
success, safety, or alignment.

\section*{Acknowledgments}

An AI coding agent assisted with manuscript preparation and artifact
verification.

\bibliographystyle{plainnat}
\bibliography{references}

\end{document}